\documentclass[traditabstract]{aa}

\usepackage{txfonts}
\usepackage{graphicx}
\usepackage{natbib}
\usepackage{longtable,dcolumn}
\usepackage{supertabular}

\begin{document}

\title{Near-infrared colours of Active Galactic Nuclei}

\author{S. Kouzuma
\and H. Yamaoka}

\institute{Graduate School of Sciences, Kyushu University, Fukuoka 812-8581, Japan \\
			 \email{kouzuma@phys.kyushu-u.ac.jp}}

\titlerunning{Near-infrared colours of AGNs}
\authorrunning{ S. Kouzuma \& H. Yamaoka}

\date{Received  / Accepted}

\abstract{
We propose near-infrared colour selection criteria to extract Active Galactic Nuclei (AGNs) 
using the near-infrared Colour-Colour Diagram (CCD) 
and predict near-infrared colour evolution with respect to redshift.  
First, we cross-identify two AGN catalogues with the 2MASS Point Source Catalogue, and 
confirm both the loci of quasars/AGNs in the near-infrared CCD and redshift-colour relations. 
In the CCD, the loci of over $70 \sim 80 \%$ of AGNs can be distinguished from the stellar locus. 
To examine the colours of quasars, 
we simulate near-infrared colours using Hyperz code. 
Assuming a realistic quasar SED, 
we derive simulated near-infrared colours of quasars with redshift (up to $z \sim 11$). 
The simulated colours can reproduce not only the redshift-colour relations 
but also the loci of quasars/AGNs in the near-infrared CCD. 
We finally discuss the possibility of contamination by other types of objects. 
We compare the locus of AGNs with the other four types of objects 
(namely, microquasars, cataclysmic variables, low mass X-ray binaries, and 
massive young stellar objects) 
which have a radiation mechanism similar to that of AGNs. In the near-infrared CCD, each type of object is located at
 a position similar to the stellar locus. 
Accordingly, it is highly probable that the four types of objects can be 
distinguished on the basis of the locus in a near-infrared CCD. 
We additionally consider contamination by distant normal galaxies. 
The near-infrared colours of several types of galaxies are also simulated using the Hyperz code. 
Although galaxies with $z\sim 1$ have near-infrared colours similar to those of AGNs, 
these galaxies are unlikely to be detected 
because they are very faint. 
In other words, few galaxies should contaminate the locus of AGNs in the near-infrared CCD. 
Consequently, we can extract reliable AGN candidates on the basis of the near-infrared CCD. 
}


\maketitle


\section{Introduction}
Active Galactic Nuclei (AGNs) are tremendous energetic sources, 
where vast amounts of energy are generated 
by gravitational accretion around supermassive black hole. 
The radiation at nearly all wavelengths enables us to detect AGNs in multiwavelength observations. 
Hence, AGNs have been studied at various wavelengths. 
Past studies show that their Spectral Energy Distributions (SEDs) are 
roughly represented by a power-law (i.e., $f_\nu \propto \nu^{-\alpha}$), 
whilst normal galaxies produce an SED that peaks at $\sim 1.6 \mu$m 
as the composite blackbody spectra of the stellar population. 
Because the colours of an object provide us with rough but essential information about its spectrum, 
colours are important clues to identify AGNs from normal stars. 

Colour selection is an efficient technique to distinguish AGNs from normal stars and 
have played an important role to extract AGN candidates without spectral observation. 
A classic method is known as the $UV$-excess \citep[$UVX$; ][]{Sandage1965-ApJ,Schmidt1983-ApJ,Boyle1990-MNRAS}. 
The $UVX$ technique exploits the fact that quasars are relatively brighter than stars at shorter wavelength 
and therefore occupy a bluer locus in a CCD with respect to stars. 
In addition, many AGN candidates have been selected on the basis of colours in various wavelengths: 
optical \citep{Richards2002-AJ}, 
optical and near-infrared \citep{Glikman2007-ApJ}, and
mid-infrared \citep{Lacy2004-ApJS,Stern2005-ApJ}. 
These studies provide us with clues about the properties of AGNs. 

Target selection of high redshift quasars has also been performed 
using their colours, mainly in optical wavelengths 
\citep[e.g., ][]{Fan2000-AJ,Fan2001-AJ,Fan2003-AJ}. 
However, near-infrared properties are required 
when we try to select targets such as higher redshift quasars, 
since the shift of the Lyman break to longer wavelengths makes 
observations difficult at optical wavelengths. 
Therefore, near-infrared selection should be useful technique to extract high-redshift quasars. 

In this paper, we present a study of the near-infrared colours of AGNs 
and demonstrate, by both observed and simulated colours, that 
the near-infrared colours can separate AGNs from normal stars. 
Additionally, we predict near-infrared colour evolution based on a Monte-Carlo simulation. 
In Sect. \ref{Data}, we introduce the catalogues of AGNs 
which are used in order to investigate the observed colours. 
We confirm the near-infrared properties of spectroscopically confirmed AGNs 
on the basis of the near-infrared CCD and 
redshift-colour relations in Sect. \ref{Properties}. 
In Sect. \ref{Simulation}, we simulate the near-infrared colours 
using Hyperz code developed by \citet{Bolzonella2000-AA} and 
demonstrate that AGNs reside in a distinct position in the near-infrared CCD.
In Sect. \ref{Discussion}, we consider the other probable objects 
which are expected to have near-infrared colours similar to those of AGNs.


\if2
\begin{figure*}[t]
\begin{tabular}{cc}
(a) & (b) \\
	\begin{minipage}[tbp]{0.5\textwidth}
	\begin{center}
		\resizebox{90mm}{!}{\includegraphics[clip]{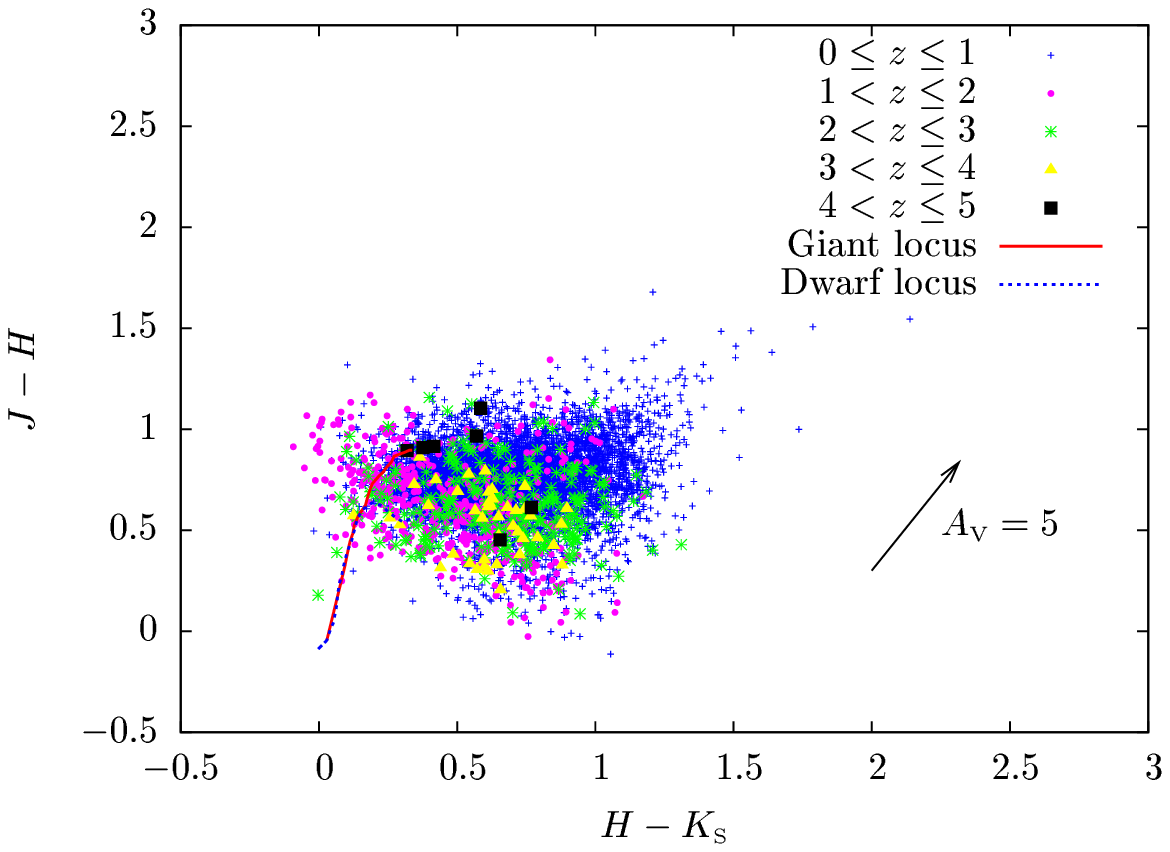}}
	\end{center}
	\end{minipage}
 &
	\begin{minipage}[htbp]{0.5\textwidth}
	\begin{center}
		\resizebox{90mm}{!}{\includegraphics[clip]{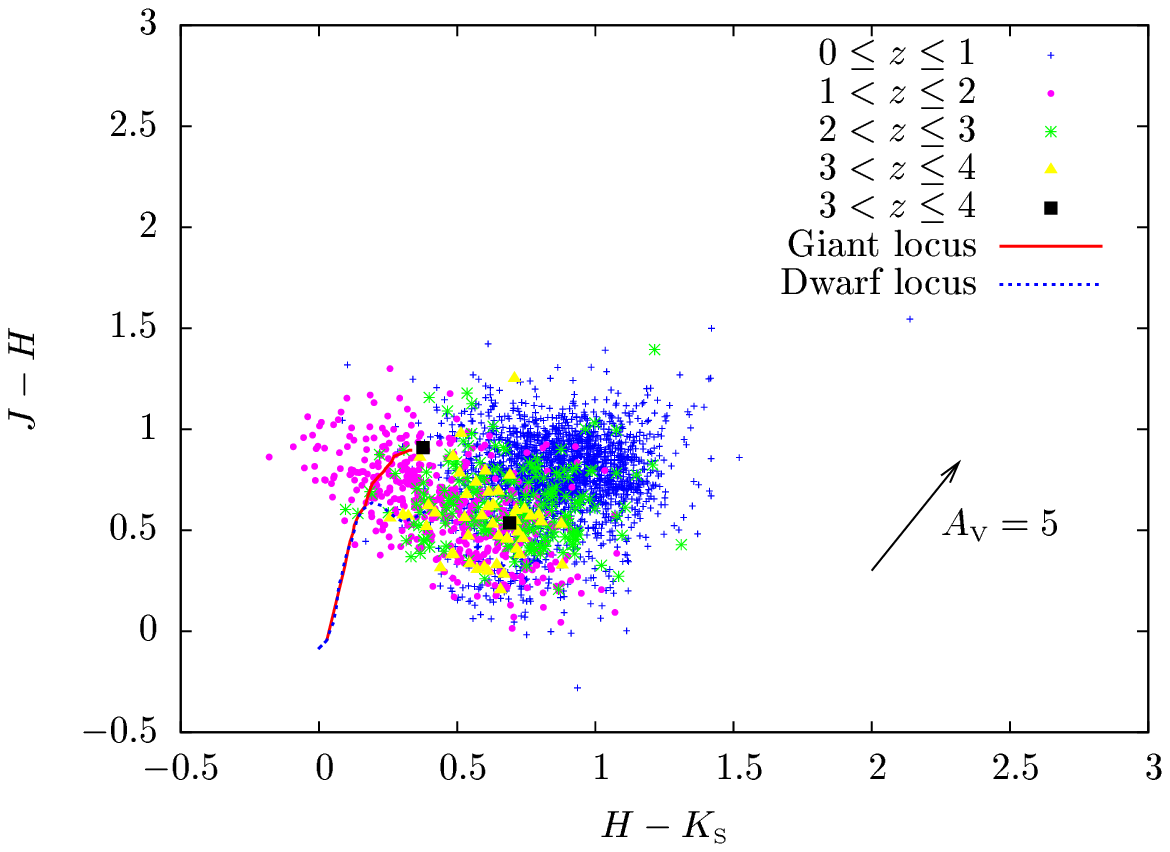}}
	\end{center}
	\end{minipage}
\end{tabular}
		\caption{(a) The distribution of AGNs in QA catalogue. 
					(b) The distribution of quasars in SQ catalogue. 
					The stellar locus in \citet{Bessell1988-PASP} and 
					the reddening vector taken from \citet{Rieke1985-ApJ} are also shown. 
		\label{CCD1}}
\end{figure*}
\fi

\begin{figure*}[tbp]
  \begin{center}
    \begin{tabular}{cc}
      (a) & (b) \\
      \resizebox{50mm}{!}{\includegraphics[]{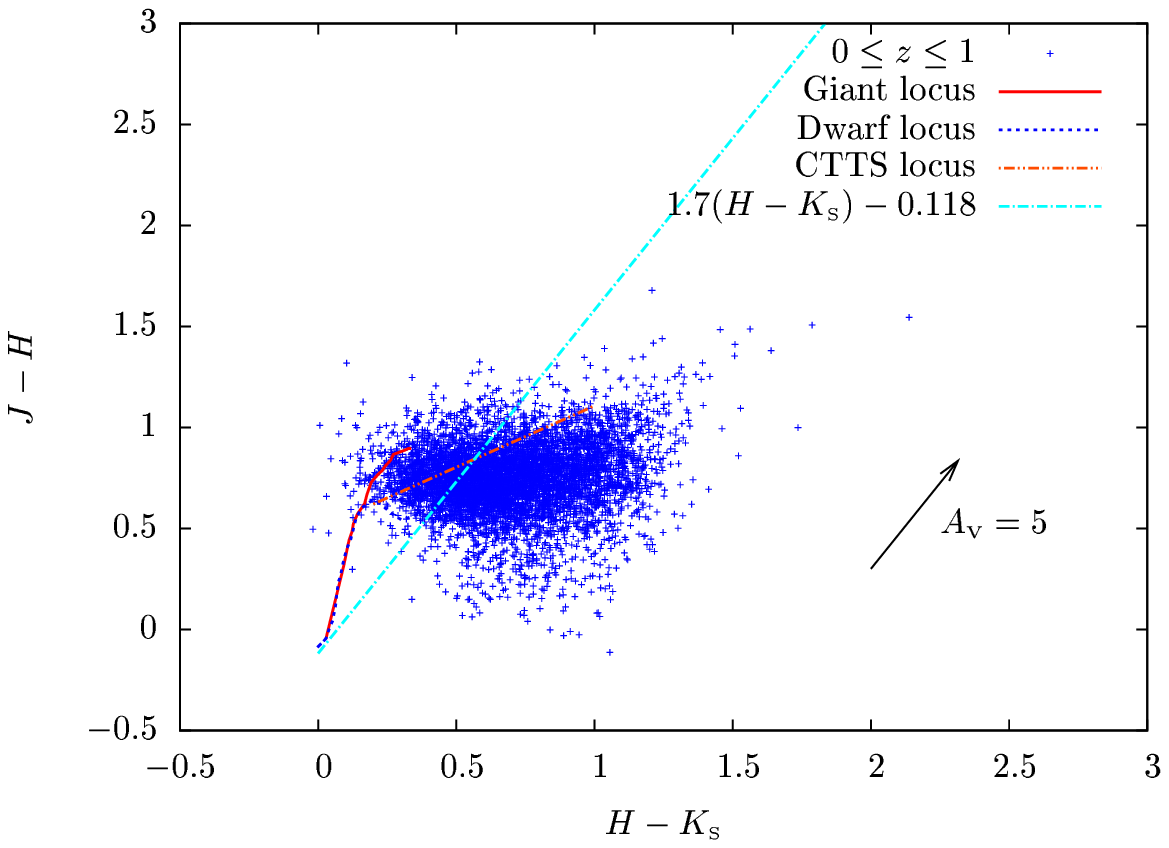}} &
      \resizebox{50mm}{!}{\includegraphics[]{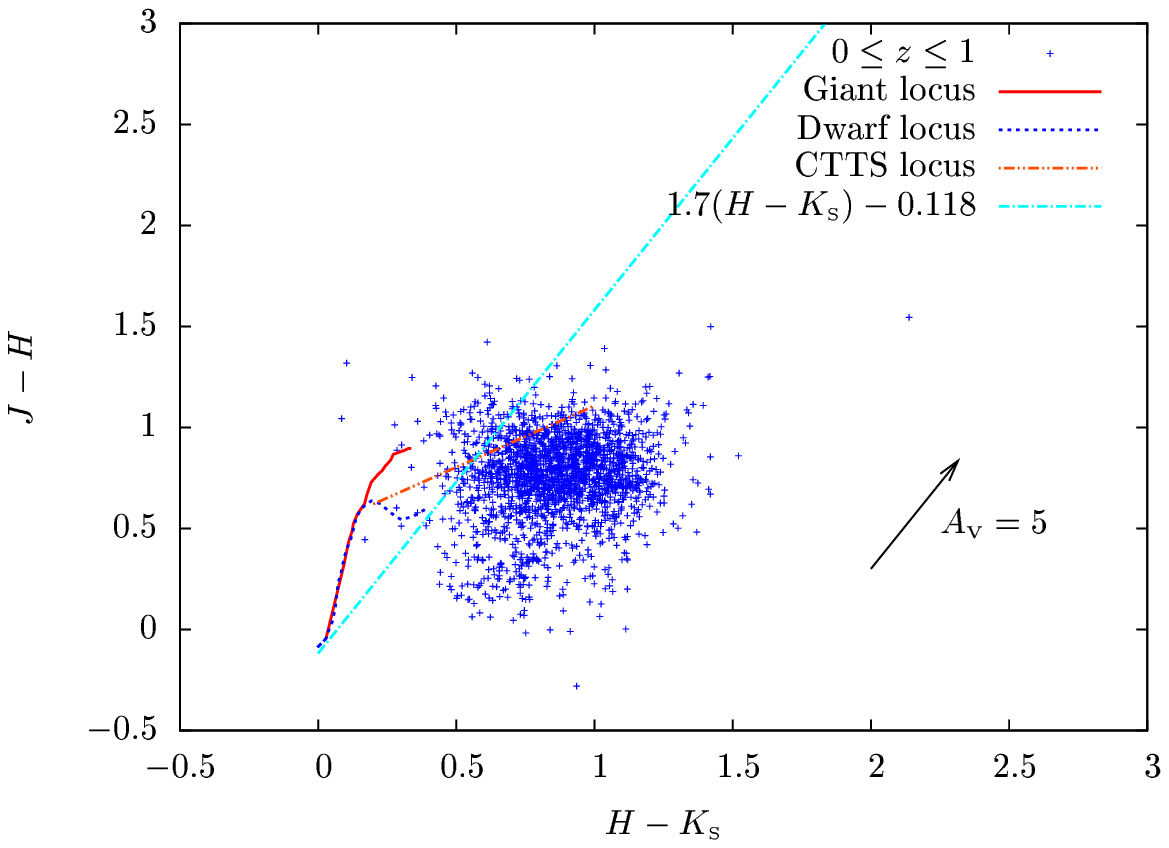}} \\
      \resizebox{50mm}{!}{\includegraphics[]{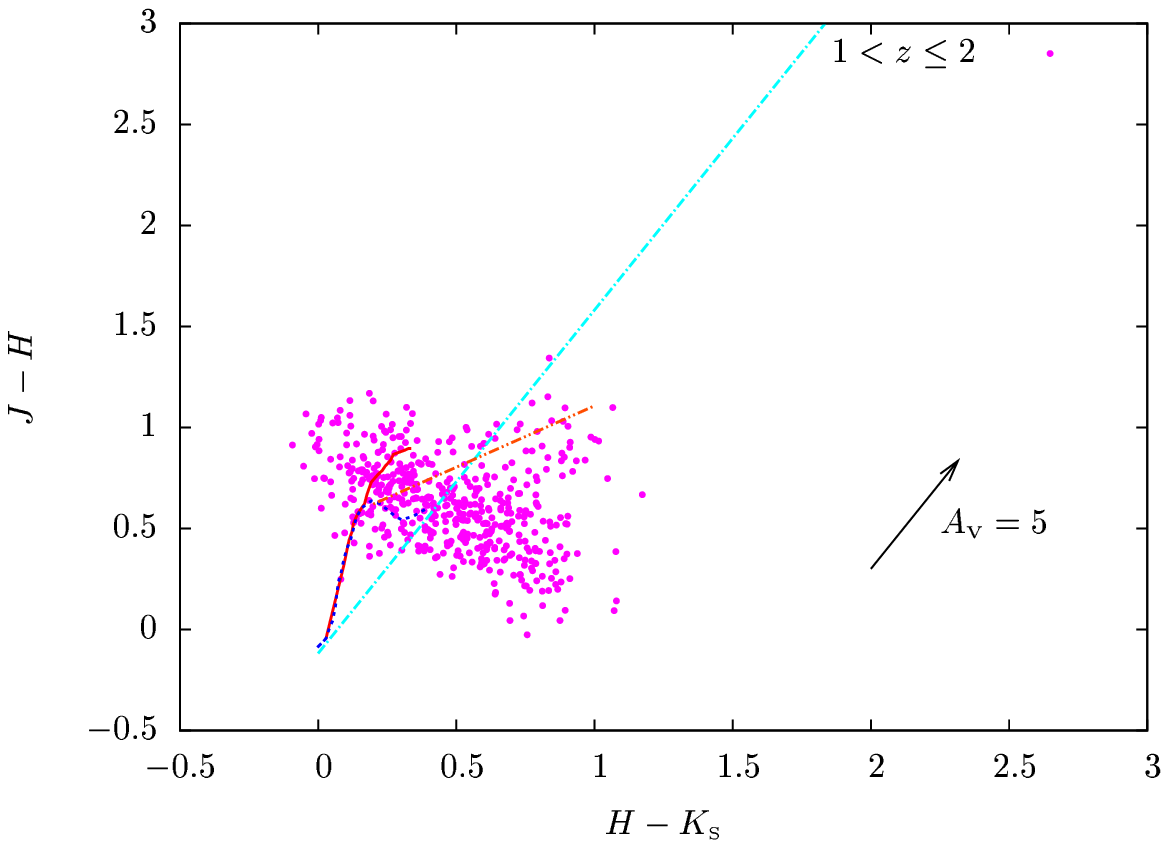}} &
      \resizebox{50mm}{!}{\includegraphics[]{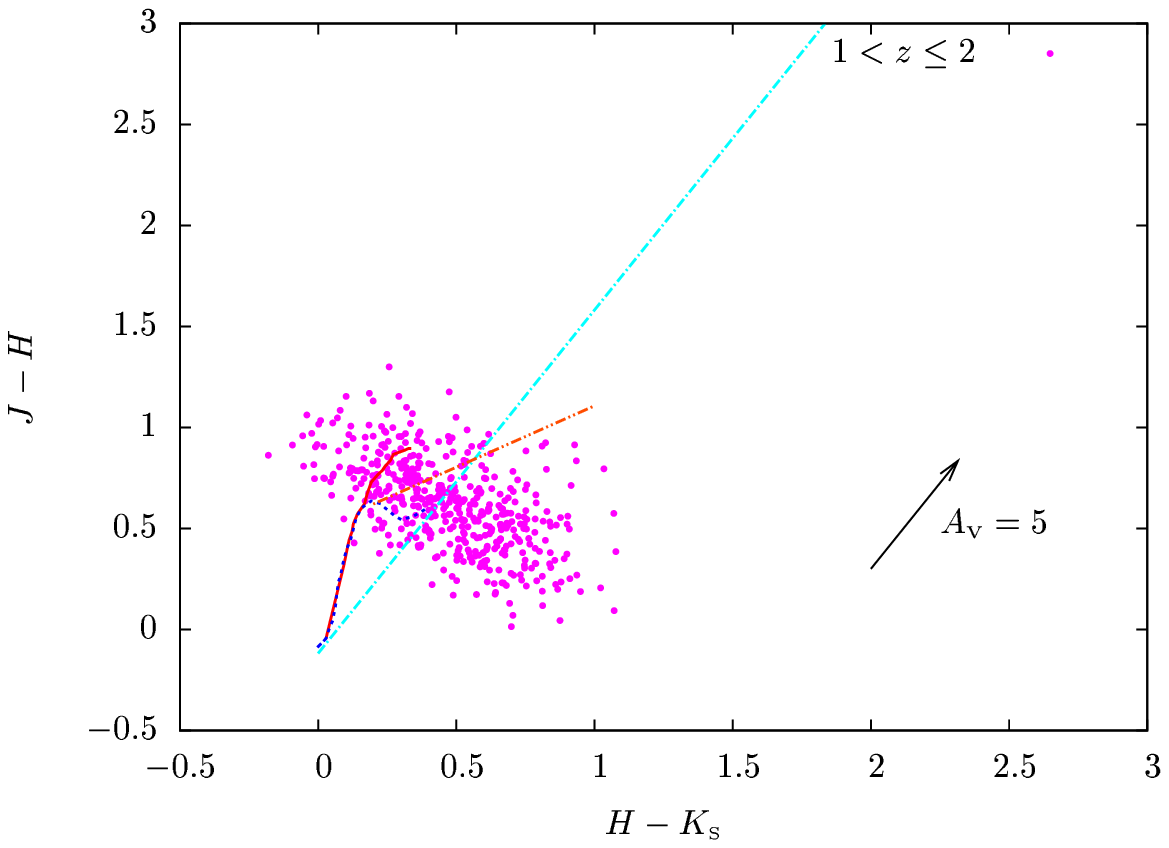}} \\
      \resizebox{50mm}{!}{\includegraphics[]{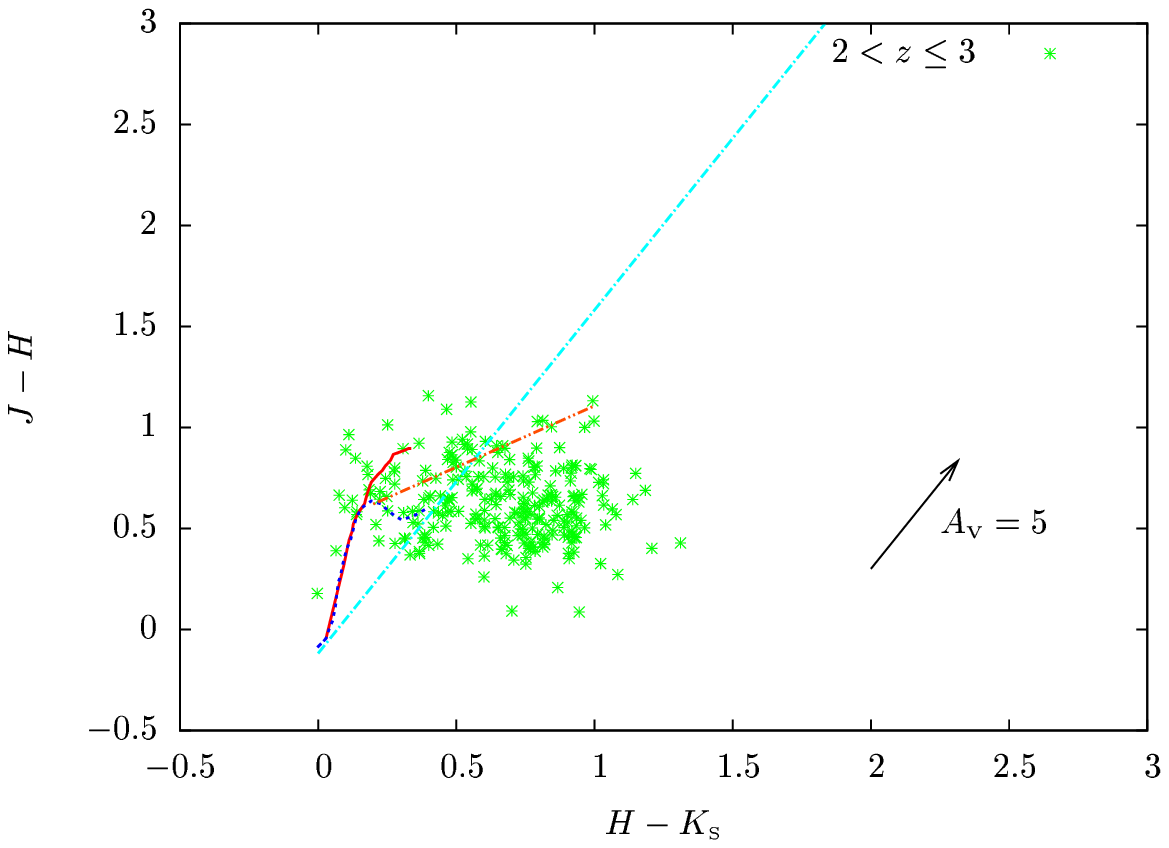}} &
      \resizebox{50mm}{!}{\includegraphics[]{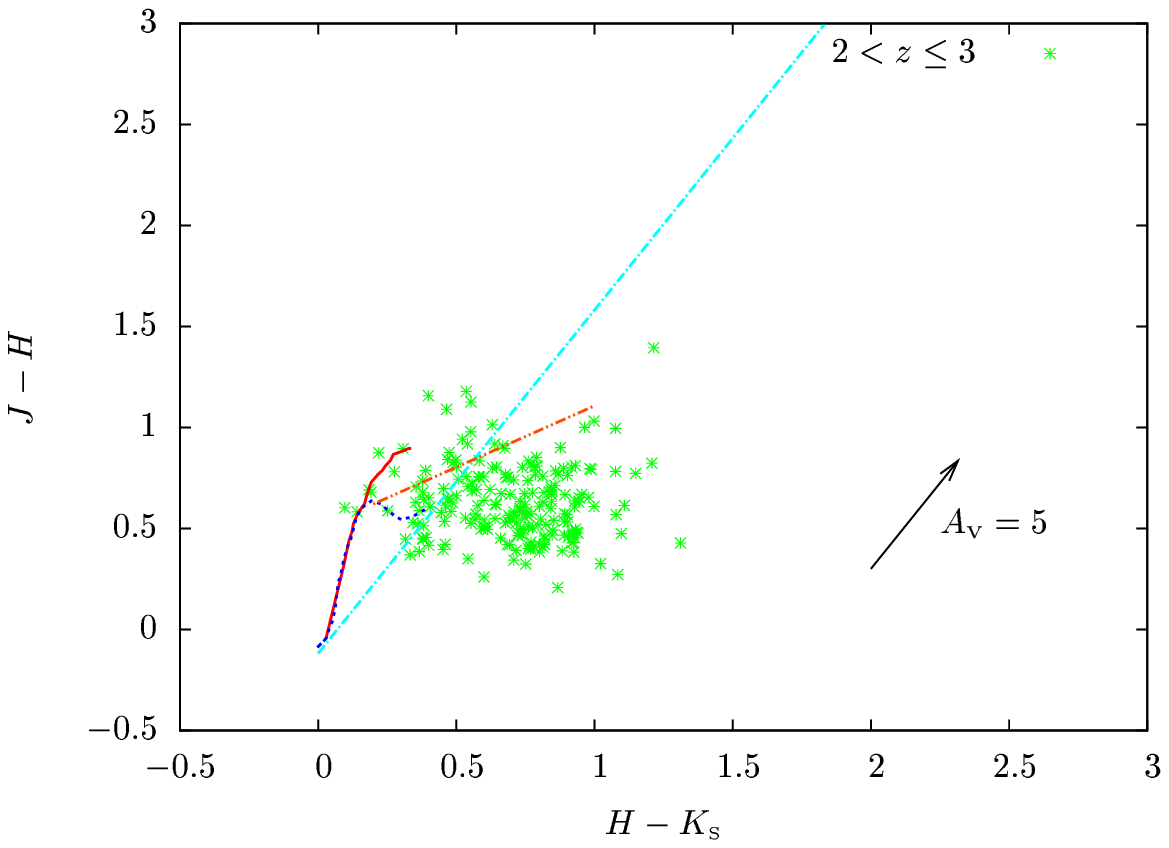}} \\
      \resizebox{50mm}{!}{\includegraphics[]{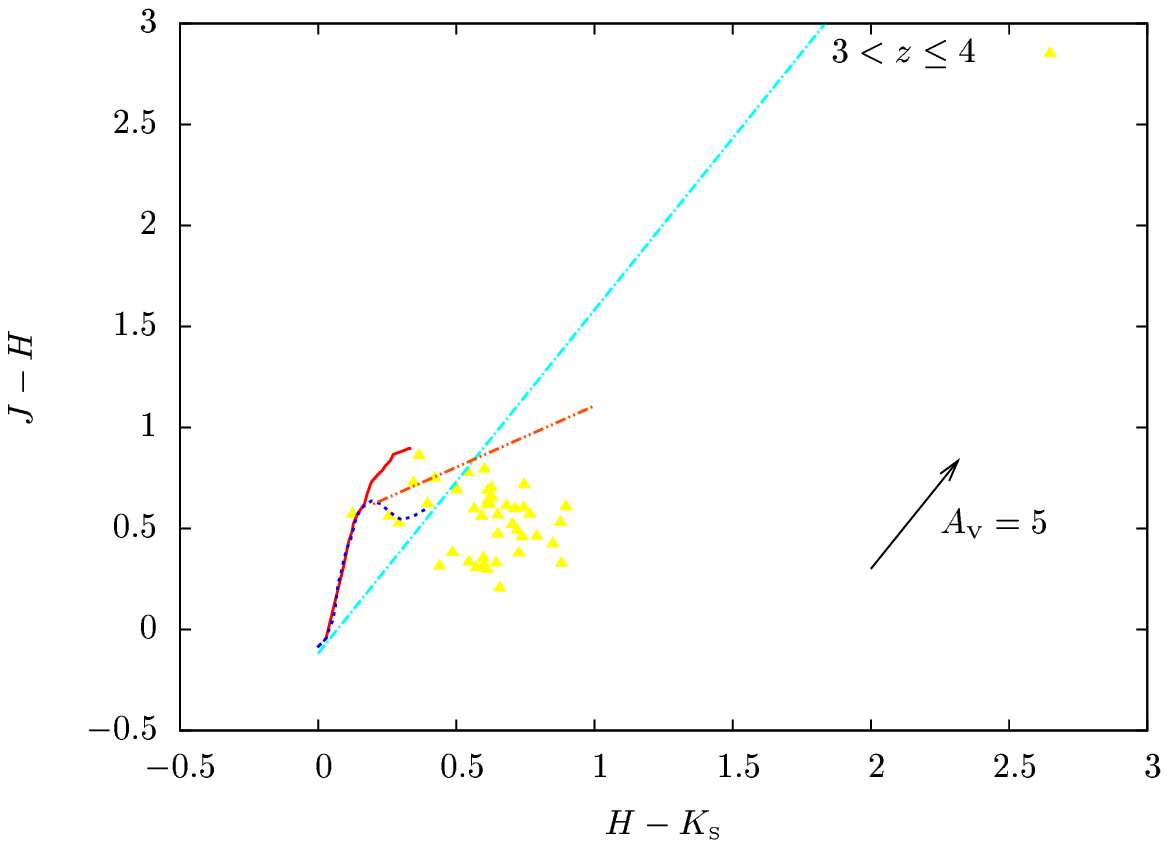}} &
      \resizebox{50mm}{!}{\includegraphics[]{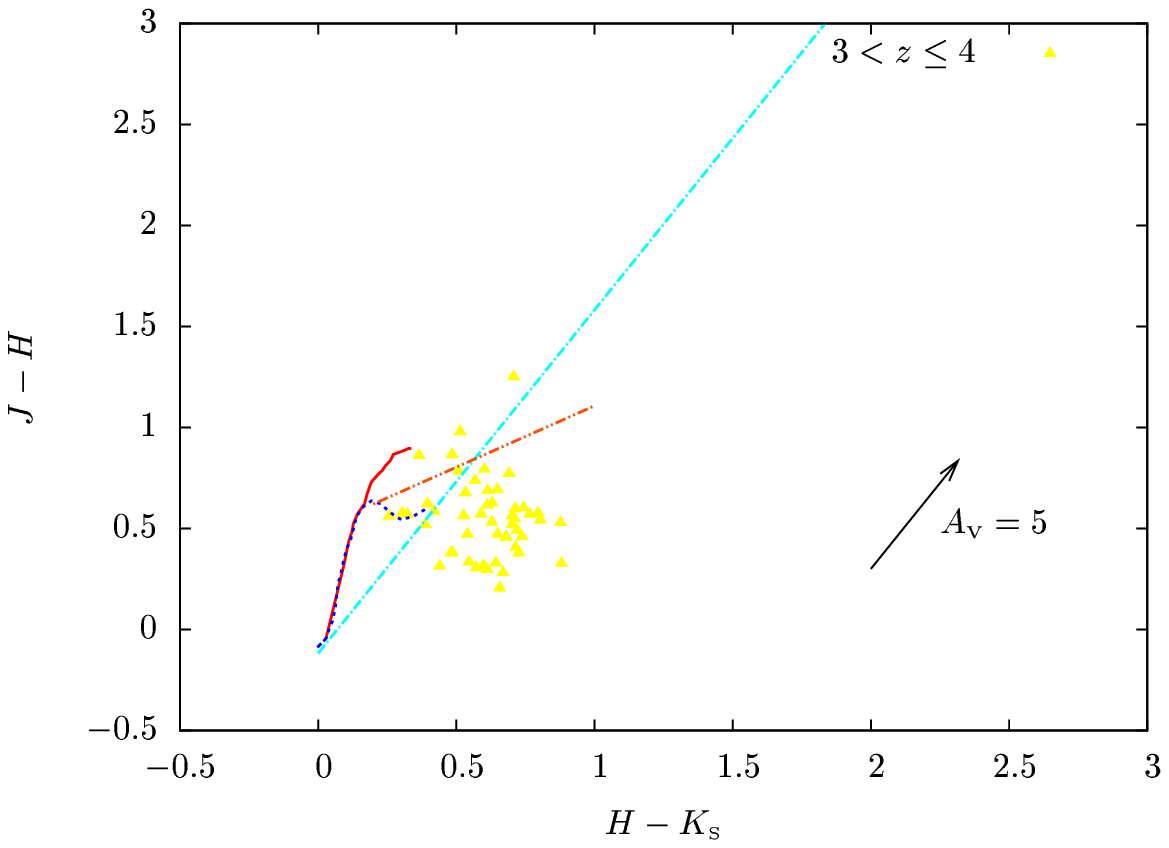}} \\
      \resizebox{50mm}{!}{\includegraphics[]{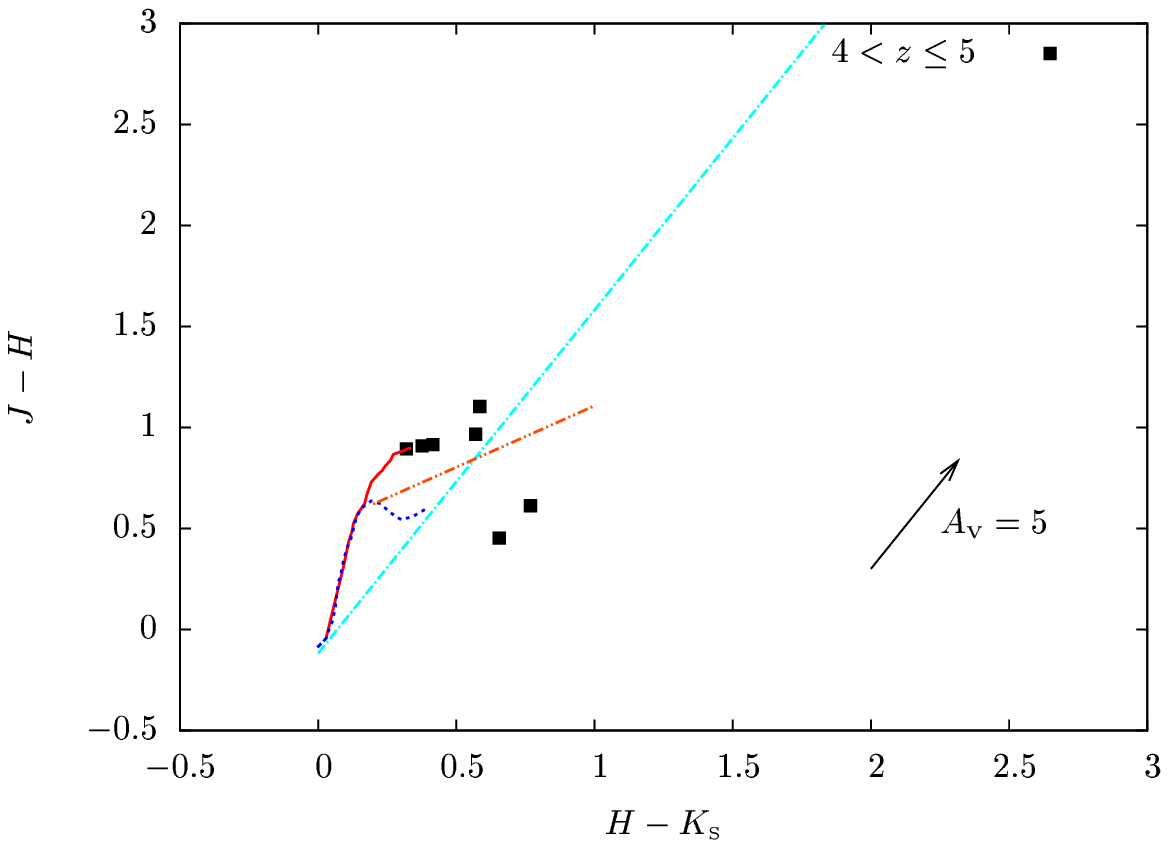}} &
      \resizebox{50mm}{!}{\includegraphics[]{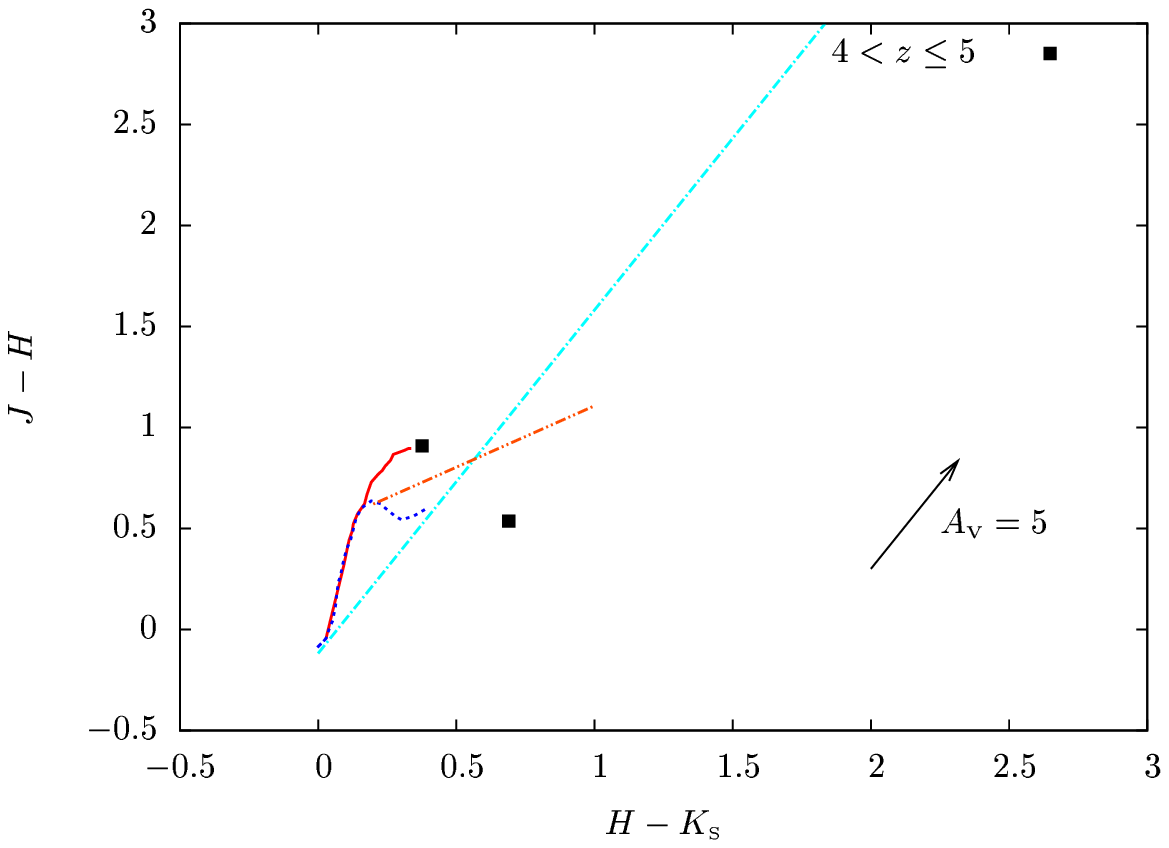}} \\
    \end{tabular}
    \caption{(a) The distribution of AGNs in the QA catalogue. 
					(b) The distribution of quasars in the SQ catalogue. 
					The stellar locus \citep{Bessell1988-PASP}, the CTTS locus \citep{Meyer1997-AJ}, and 
					the reddening vector taken from \citet{Rieke1985-ApJ} are also shown. 
		\label{CCD1}}
  \end{center}
\end{figure*}

\section{Data}\label{Data}

We examine the near-infrared properties of quasars/AGNs using 2MASS magnitudes. 
The samples of quasars/AGNs are extracted from 
Sloan Digital Sky Survey Data Release 5 (SDSS-DR5) quasar catalog and 
Quasars and Active Galactic Nuclei (12th Ed. ) 
and these catalogues are briefly introduced below. 

\subsection{2MASS}
The Two Micron All Sky Survey \citep[2MASS
\footnote{2MASS web site (http://www.ipac.caltech.edu/2mass/)};  ][]{Skrutskie2006-AJ}
is a project which observed 99.998\% of the whole sky 
at the J (1.25 $\mu$m), H (1.65 $\mu$m), Ks (2.16 $\mu$m) bands, 
at Mt. Hopkins, AZ (the Northern Hemisphere) and at CTIO, Chile (the Southern Hemisphere)
between June 1997 and February 2001. 
The instruments are both highly automated 1.3-m telescopes equipped with three-channel cameras, 
each channel consisting of a 256 $\times$ 256 array of HgCdTe detectors. 
The 2MASS obtained 4 121 439 FITS images (pixel size $\sim2''_{\cdot}0$) with 7.8 s of integration time. 
The $10 \sigma$ point-source detection levels are better than 15.8, 15.1, and 14.3 mag 
at J, H, and K$_\textnormal{\tiny S}$ bands. 
The Point Source Catalogue (PSC) was produced using these images and catalogued 470 992 970 sources. 
In the 2MASS web site, the images and the PSC are open to the public and are easily available.

\begin{table*}[tbp]
\begin{center}
\begin{tabular}{crrrrrr}
\hline \hline
redshift & \multicolumn{3}{c}{QA catalogue} & \multicolumn{3}{c}{SQ catalogue} \\ 
range & \multicolumn{3}{c}{-----------------------------------} & \multicolumn{3}{c}{-----------------------------------} \\ 
& Region I & Region II & total & Region I & Region II & total \\ \hline
$0 \leq z \leq 1$ & 1 671 (27) & 4 480 (73) & 6 151 & 222 (11) & 1 869 (89) & 2 091 \\
$1 < z \leq 2$ & 238 (47) & 265 (53) & 503 & 222 (47) & 249 (53) & 471 \\
$2 < z \leq 3$ & 67 (25) & 197 (75) & 264 & 38 (19) & 165 (81) & 203 \\
$3 < z \leq 4$ & 7 (16) & 36 (84) & 43 & 9 (18) & 41 (82) & 50 \\
$4 < z \leq 5$ & 5 (71) & 2 (29) & 7 & 1 (50) & 1 (50) & 2 \\ \hline
total & 1 998 (29) & 4 970 (71) & 6 968 & 500 (18) & 2 317 (82) & 2 817 \\ \hline
\end{tabular}
\caption{
The number of objects distributed in the Region I and II. 
Of 7 061 AGNs detected in the QA catalogue, 
93 do not have a measured redshift. The values in parentheses represent percentages of quasars/AGNs residing in each region. 
\label{Ratio}}
\end{center}
\end{table*}

\subsection{SDSS-DR5 quasar catalog}
The Sloan Digital Sky Survey (SDSS) 
provides a photometrically and astrometrically 
calibrated digital imaging survey of $\pi$ sr above Galactic latitude $30^\circ$ 
in five broad optical bands to a depth of $g' \sim 23$ mag \citep{York2000-AJ}. 
Many astronomical catalogues have been produced by this survey. 

The SDSS quasar catalog IV \citep[][hereafter SQ]{Schneider2007-AJ} is 
the forth edition of the SDSS quasar catalog I \citep{Schneider2002-AJ}, 
which is made from the SDSS Fifth Data Release \citep{Adelman2007-ApJS}. 
The SQ catalogue consists of 77 429 quasars, 
the vast majority of which were discovered by the SDSS. 
The area covered by the catalogue is $\approx 5740$ deg$^2$. 
The quasar redshifts range from 0.08 to 5.41, with a median value of 1.48. 
The positional accuracy of each object is better than $0_\cdot''2$.

\subsection{Quasars and Active Galactic Nuclei (12th Ed.)}
The catalogue of Quasars and Active Galactic Nuclei (12th Ed.) \citep[][hereafter QA]{Veron2006-AA} is 
the 12th edition of the catalogue of quasars first published in 1971 by De Veny et al.. The QA catalogue contains 85 221 quasars, 1 122 BL Lac objects and 
21 737 active galaxies (including 9 628 Seyfert 1). 
this catalogue includes position and redshift as well as optical brightness (U, B, V) and 6 cm and 20 cm flux densities when available. 
The positional accuracy is better than $1_\cdot''0$.

\section{Near-infrared properties of AGNs}\label{Properties}

\subsection{Extraction of Near-infrared counterpart}
The sources in two of the above-mentioned AGN catalogues (SQ and QA) 
were cross-identified with 2MASS PSC, 
and we extracted a near-infrared counterpart of each source. 
As mentioned in the previous section, 
the positional accuracies in both catalogues are better than $1''$. 
Therefore, we identified an near-infrared counterpart 
when a 2MASS source is located within $1''$ of a SQ/QA position. 

As a result of the extraction, 
we have derived 9 658 (SQ catalogue) and 14 078 (QA catalogue) 
near-infrared counterparts. 
For investigating the near-infrared properties using 2MASS magnitudes, 
we used only 2 817 (SQ) and 7 061 (QA) objects where 2MASS photometric quality flags are 
better than B (signal-to-noise ratio (S/N) $>7\sigma$).

\subsection{Colour-colour diagram}

Near-infrared ($H-K_\textnormal{\tiny S}$)-($J-H$) CCD is 
a powerful tool to investigate the property of celestial objects. 
We investigated the near-infrared properties of quasars/AGNs using near-infrared CCD. 

Figure \ref{CCD1} shows the distributions of quasars/AGNs in a ($H-K_\textnormal{\tiny S}$)-($J-H$) CCD. 
In previous studies, the intrinsic loci of stars and Classical T Tauri Stars (CTTS) 
were well defined by \citet{Bessell1988-PASP} and \citet{Meyer1997-AJ}. 
Their loci are also shown in the CCD. 
\citet{Bessell1988-PASP} and the Caltech (CIT) systems are transformed into the 2MASS photometric system 
by the method introduced by \citet{Carpenter2001-AJ}.  
The reddening vector, taken from \citet{Rieke1985-ApJ}, is also shown in the diagram. 
Because the stellar and CTTS loci can only shift along the reddening vector, 
most of these types of stars fundamentally should not be located in the region described by the following equations. 
\begin{eqnarray}
(J-H)  \leq 1.70(H-K_\textnormal{\tiny S})-0.118 \label{Star}\\
(J-H)  \leq 0.61 (H-K_\textnormal{\tiny S})+0.50 \label{CTTS}
\end{eqnarray}
Equation (\ref{Star}) represents the lower limit line where normal stars can reside and 
Equation (\ref{CTTS}) represents the lower limit line where CTTS can reside. 
Both lines are also shown in Fig. \ref{CCD1}. 
Below, we call the region enclosed by Equations (\ref{Star}) and (\ref{CTTS}) ``Region II'' 
and all other regions ``Region I''. 
In Fig. \ref{CCD1}, we can see that most of the quasar/AGNs 
are located in clearly different areas than the stellar loci.
The distributions of the quasar/AGNs are on the right side of the stellar loci in the CCD, 
i.e., they have a $(J-H)$ colour similar to that of normal stars 
but have a $(H-K_\textnormal{\tiny S})$ colour redder than that of normal stars.  
Table \ref{Ratio} counts the number of objects in each region. 
It shows that 70\% of AGNs and 80\% of quasars are distributed in Region II. 
Hence, the near-infrared selection of quasars can be more effective than that of other types of AGN. 
Especially, $\sim 90\%$ of low redshift quasars with $0 \leq z \leq 1$ reside in Region II, 
so these quasars are rarely missed. 
However, objects with $1< z \leq 2$ or $4<z \leq 5$ tend to have a bluer colour in $(H-K_\textnormal{\tiny S})$ 
than objects with other redshift ranges, which is similar to the colour of normal stars. 
Therefore, some of these quasars/AGNs might be missed. 
The difference of the loci between quasars/AGNs and normal stars is 
probably due to the difference of the radiation mechanism 
because the dominant radiation of quasars/AGNs is not blackbody radiation. 

This colour property is considered to be caused by K-excess \citep{Warren2000-MNRAS}. 
They proposed a $KX$ method where quasars with a ($V-J$) colour similar to 
that of stars would be redder in ($J-K$) colour. 
In other words, this $KX$ method can separate quasars and stars on the basis of their colours. 
This technique has been used for selecting quasar candidates 
\citep[e.g., ][]{Smail2008-MNRAS,Jurek2008-MNRAS,Nakos2009-AA}.
The present work is a variant of the original $KX$ technique, 
using the $(J-H)$ versus $(H-K_\textnormal{\tiny S})$ diagram. 


\subsection{Colours versus redshift}

\begin{figure}[tbp]
	\begin{center}
		\resizebox{90mm}{!}{\includegraphics[clip]{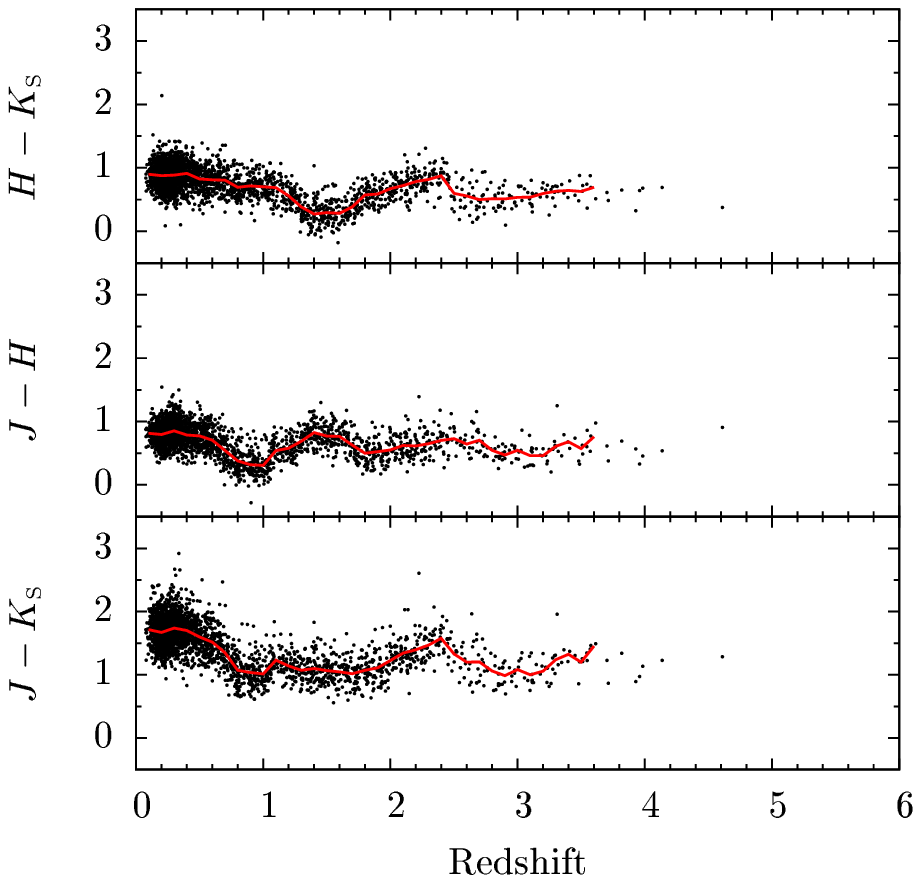}}
	\end{center}
		\caption{Colours versus redshift for SDSS quasars. 
		         The redshifts are taken from the SQ catalogue. 
				 The red solid lines show the average colour evolutions with respect to redshift. 
		\label{Z-Colours}}
\end{figure}

In Fig. \ref{Z-Colours}, we plot the SDSS quasars, in three colours versus redshift 
together with average colour evolutions with respect to redshift. 
The redshifts are taken from the SQ catalogue. 

Each colour undergoes only a small change or dispersion with redshift. 
This is probably due to a variety of spectral shapes and/or a variety of extinctions. 
In the near-infrared CCD, this small colour change causes a small difference of AGN locus. 

These properties can be reproduced by the simulation as mentioned below.

\section{Simulating the near-infrared colours of quasars}\label{Simulation}
\if2
\begin{figure}[tbp]
	\begin{center}
		\resizebox{90mm}{!}{\includegraphics[clip]{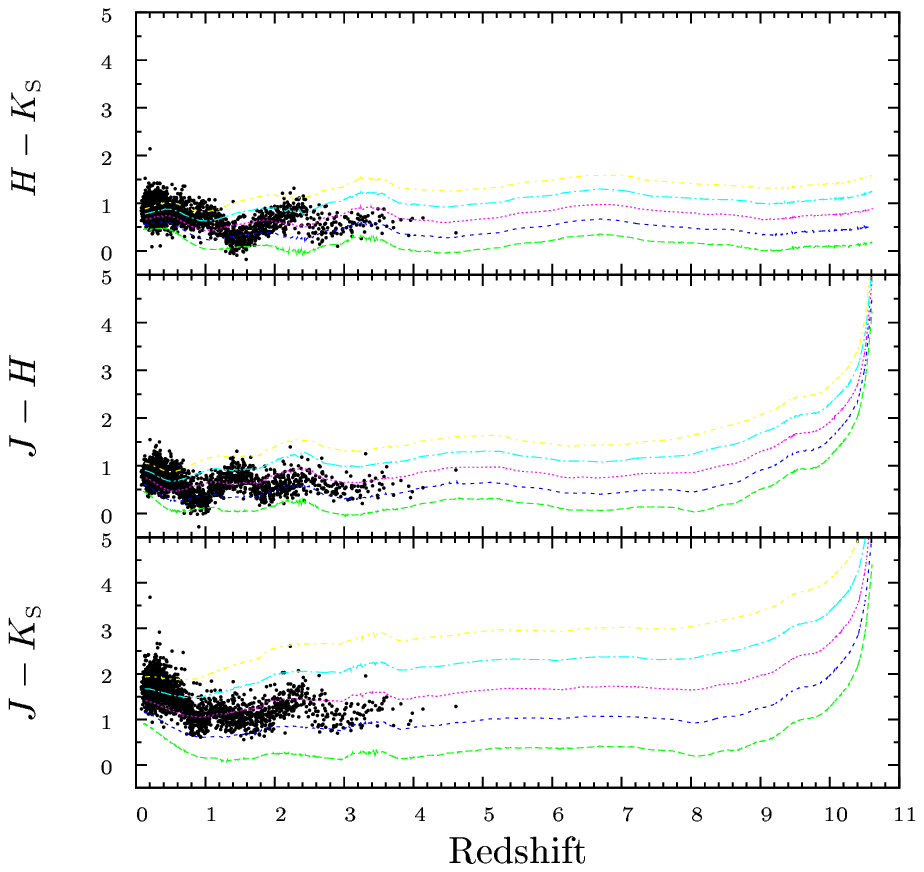}}
	\end{center}
		\caption{Colours versus redshift for SDSS quasars.   
		\label{}}
\end{figure}
\fi

\begin{figure*}[tbp]
	\begin{center}
		\resizebox{90mm}{!}{\includegraphics[clip]{figure/z-color-simulation.eps}}
		\resizebox{90mm}{!}{\includegraphics[clip]{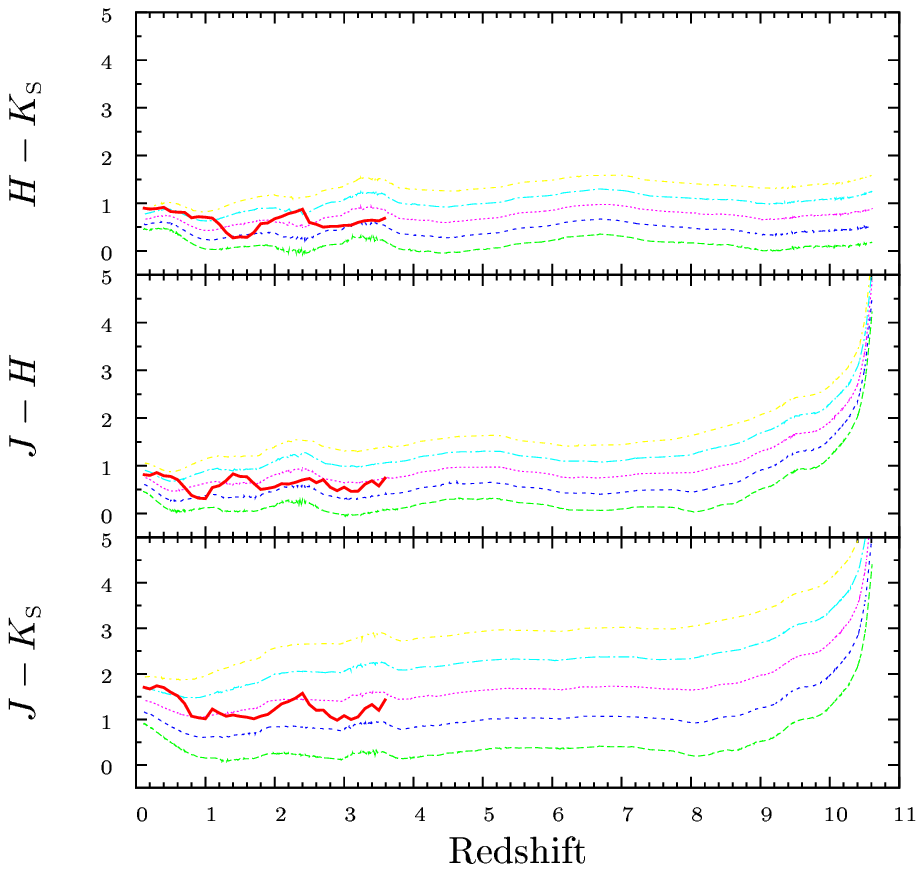}}
		\caption{Simulated colours versus redshift. 
				 The curves represent the simulated colour evolutions 
					with $A_\textnormal{\tiny V}=0,1,2,3,4$ (from bottom to top), respectively. 
				 The SDSS quasars (left panel) and the average colour evolution (right panel) 
					shown in Fig. \ref{Z-Colours} are also plotted in the diagram. 
		\label{Z-Simulated-Colours}}
	\end{center}
\end{figure*}

\begin{figure}[tbp]
	\begin{center}
		\resizebox{90mm}{!}{\includegraphics[clip]{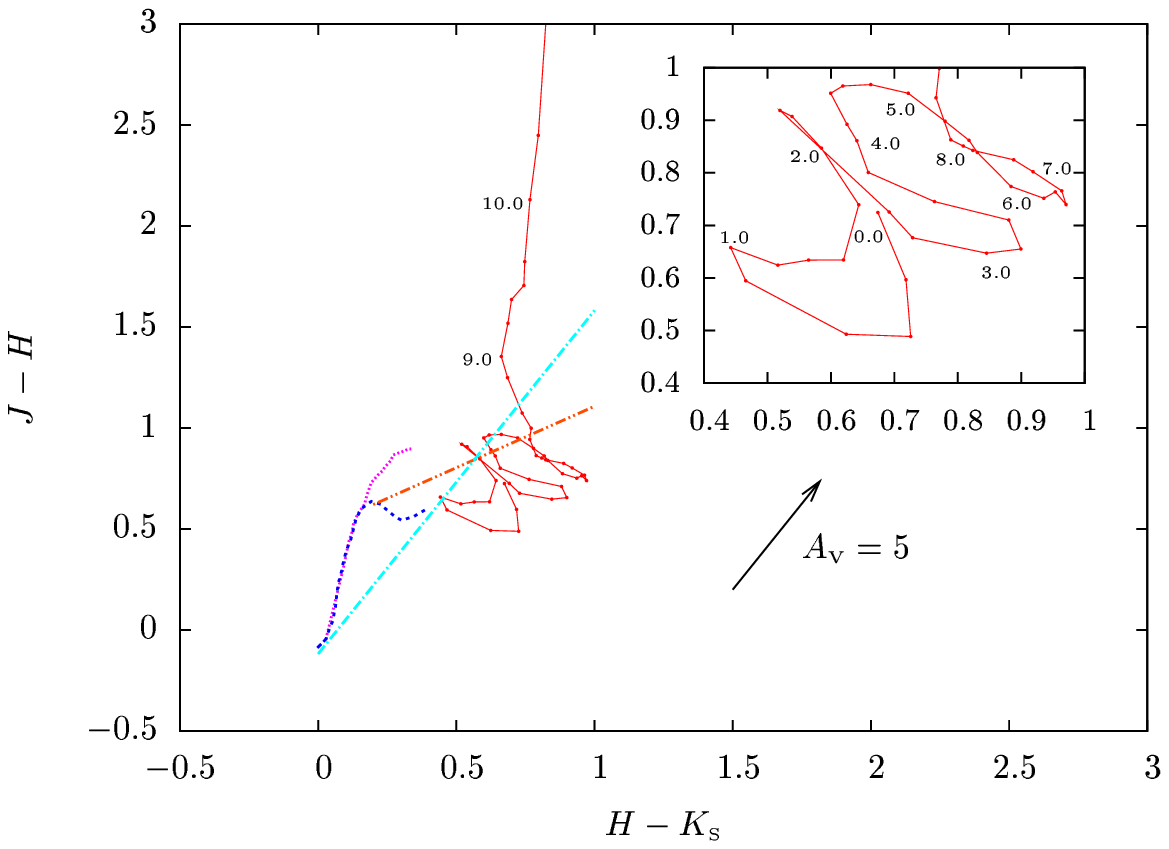}}
	\end{center}
		\caption{Simulated colour evolution with respect to redshift in the ($H-K_\textnormal{\tiny S}$)-($J-H$) diagram. 
				The stellar locus and the reddening vector are also shown in the diagram, 
					which are same as in Fig. \ref{CCD1}	. 
		\label{CCD-Simulation}}
\end{figure}

\begin{table}[tbp]
\begin{center}
\begin{tabular}{crrr}
\hline \hline
$A_\textnormal{\tiny V}$ & $J-K_\textnormal{\tiny S}$ & $J-H$ & $H-K_\textnormal{\tiny S}$ \\ \hline
0 & 0.78 (0) & 0.94 (0) & 0.86 (0) \\
1 & 0.94 (0) & 0.69 (0) & 0.60 (0) \\
2 & 0.26 (17) & 0.29 (9) & 0.14 (84) \\
3 & 0.79 (0) & 0.82 (0) & 0.59 (0) \\
4 & 1.0 (0) & 1.0 (0) & 0.88 (0) \\ \hline
\end{tabular}
\caption{
Results of a KS test between average colour evolution and simulated colour evolution. 
The decimal values represent KS distance between two data and 
the values in parentheses represent significance level (percentage) for each KS test. 
\label{KS-test}}
\end{center}
\end{table}

In this section, 
we demonstrate that the locus of quasars is well separated from that of normal stars 
on the basis of a simulation using a realistic SED of quasars. 

In order to simulate the near-infrared colours of quasars, 
we performed a Monte-Carlo simulation with Hyperz code \citep{Bolzonella2000-AA}. 
The Hyperz code can calculate photometric redshift based on an inputted spectral template library, 
which finds the best fit SED by minimizing the $\chi^2$ derived from 
the comparison among the observed SED and expected SEDs. 
The reddening effects are taken into account according to a selected reddening law. 
Although this code is usually used for estimating photometric redshifts, 
we use it to derive the near-infrared colours at various redshifts. 

First, we made a magnitude list containing randomly generated J, H, K$_\textnormal{\tiny S}$ magnitudes, 
ranging from 8 to 16 mag (roughly coincident with a reliable range of 2MASS magnitude) 
and produced 100 000 data sets. 
These data sets were subjected to 
SED fittings using the Hyperz code. 
A realistic SED of quasars was taken from \citet{Polletta2007-ApJ} 
(i.e., QSO1 in \citet{Polletta2007-ApJ} are used). 
According to \citet{Polletta2007-ApJ}, 
the SED of the QSO1 is derived by combining the SDSS quasar composite spectrum 
and rest-frame IR data of a sample of 3 SDSS/SWIRE quasars \citep{Hatziminaoglou2005-AJ}. 
We used the reddening law from \citet{Calzetti2000-ApJ}, which is prepared by default in Hyperz code. Inputting the data sets into the Hyperz code, 
we derived photometric redshifts with the probabilities associated with the value of $\chi^2$. 
We only used objects having probabilities of $\geq 99\%$. 

Figure \ref{Z-Simulated-Colours} shows the simulated colour evolutions with respect to redshift. 
The curves in each diagram represent the simulated colours 
with $A_\textnormal{\tiny V} =0$, 1, 2, 3, 4 (from bottom to top), respectively. 
To find the best fits for the average colour curves, 
we performed Kolmogorov-Smirnov (KS) tests between average colour curves and each simulated colour curve. 
Table \ref{KS-test} shows the result of the KS tests. 
In all three colours, the colour evolution with $A_\textnormal{\tiny V}=2$ is
 the best fit among five $A_\textnormal{\tiny V}$ values. 
In addition, the redshift-colour relations of SQ quasars can be roughly reproduced 
by simulated curves with $0\lesssim A_\textnormal{\tiny V} \lesssim 3$. 
A variety of extinctions probably generate the dispersion of the colours. 
It should be noted that both ($J-H$) and ($J-K_\textnormal{\tiny S}$) colours steeply 
increase over $z\sim 9$. 
This is due to shifting the Lyman break over the J-band wavelength. 
This property can be useful for extracting high-redshift quasars. 

In Fig. \ref{CCD-Simulation}, the simulated colours with $A_\textnormal{\tiny V}=2$ are shown 
in the ($H-K_\textnormal{\tiny S}$)-($J-H$) CCD, tracked by redshift evolution. 
An important point is that the simulated position is well separated from the stellar locus, 
that is, it is consistent with the loci of quasars/AGNs shown in Fig. \ref{CCD1}. 
A variety of extinctions causes the dispersion of the simulated position and 
this can probably reproduce the dispersion of the loci of quasars/AGNs in Fig. \ref{CCD1}. 
It is also consistent with the fact that 
the quasars with $0 \leq z \leq 1$ have relatively redder colour in ($H-K_\textnormal{\tiny S}$) compared with 
the quasars with $1 \leq z \leq$ 2. 

Although it is difficult to distinguish high-redshift quasars among $z\lesssim 8$,
 we can extract high-redshift quasar candidates with $z \gtrsim 8$ 
on the basis of a ($H-K_\textnormal{\tiny S}$)-($J-H$) diagram 
because the ($J-H$) colour steeply increases over $z \sim 8$.

\section{Discussion}\label{Discussion}
\subsection{Other probable objects} \label{Probabilities}

\begin{figure*}[tbp]
\begin{tabular}{cc}
(a) & (b) \\
	\begin{minipage}[htbp]{0.5\textwidth}
	\begin{center}
		\resizebox{80mm}{!}{\includegraphics[clip]{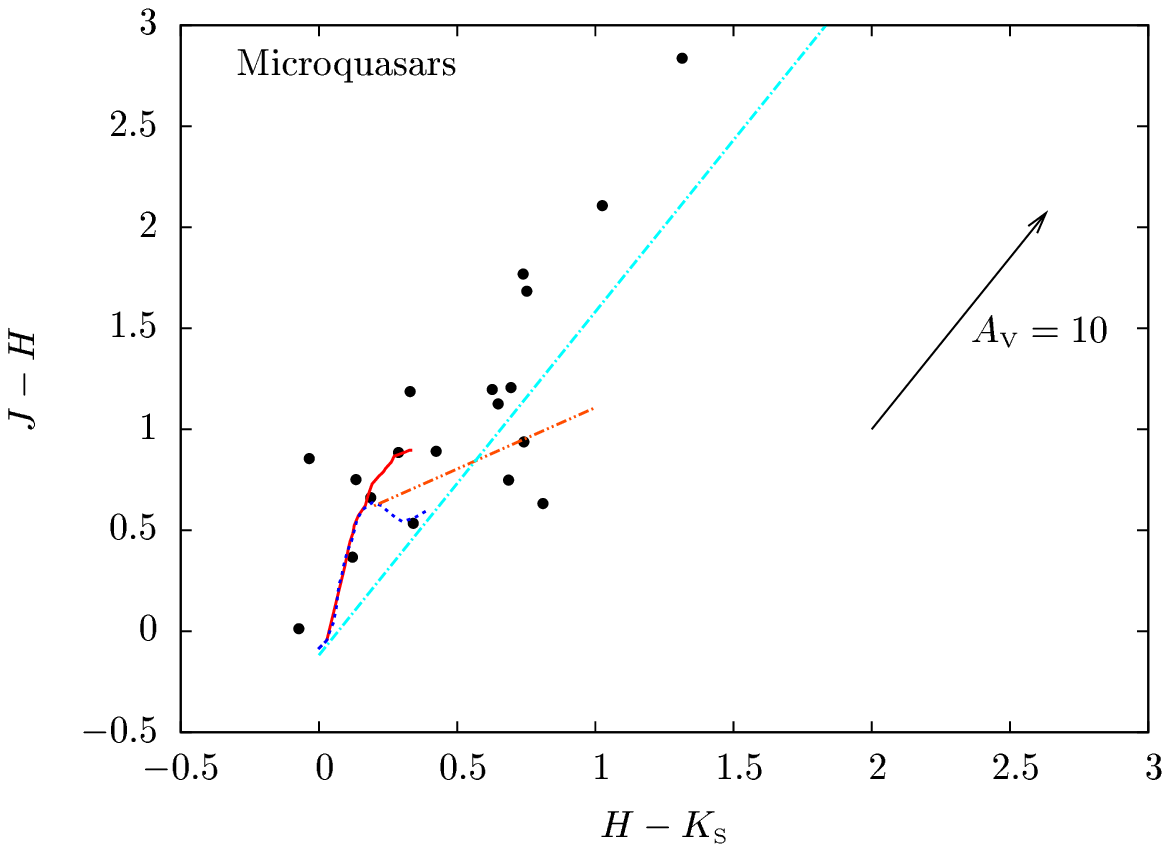}}
	\end{center}
	\end{minipage}
 &	
\begin{minipage}[htbp]{0.5\textwidth}
	\begin{center}
		\resizebox{80mm}{!}{\includegraphics[clip]{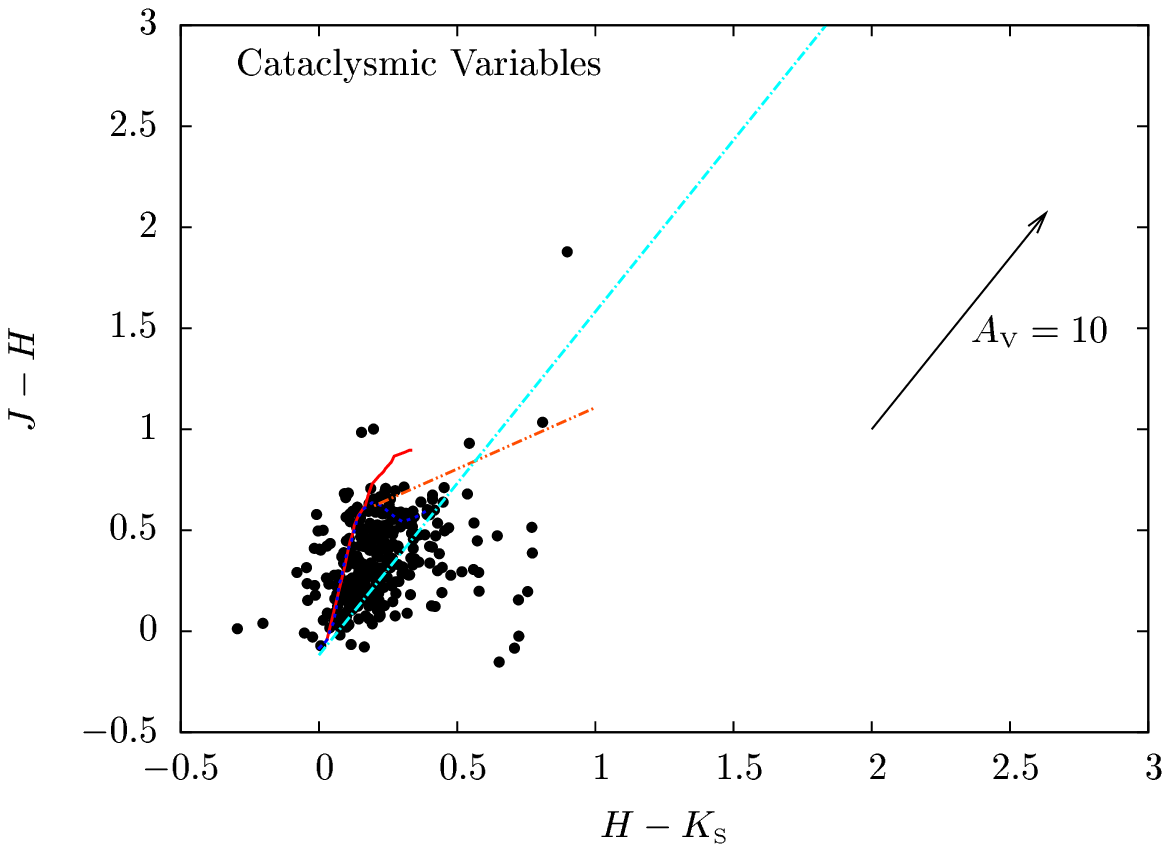}}
	\end{center}
	\end{minipage} \\
(c) & (d) \\
	\begin{minipage}[htbp]{0.5\textwidth}
	\begin{center}
		\resizebox{80mm}{!}{\includegraphics[clip]{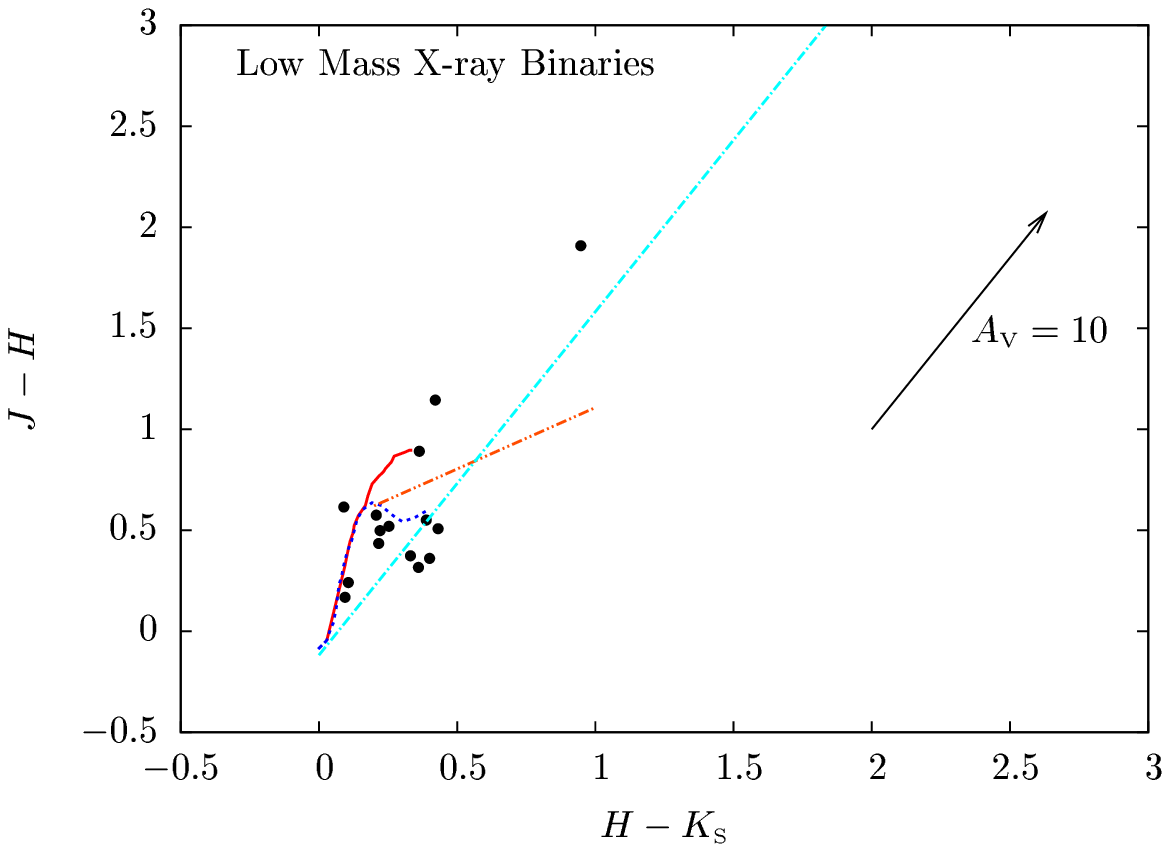}}
	\end{center}
	\end{minipage}
 &	
\begin{minipage}[htbp]{0.5\textwidth}
	\begin{center}
		\resizebox{80mm}{!}{\includegraphics[clip]{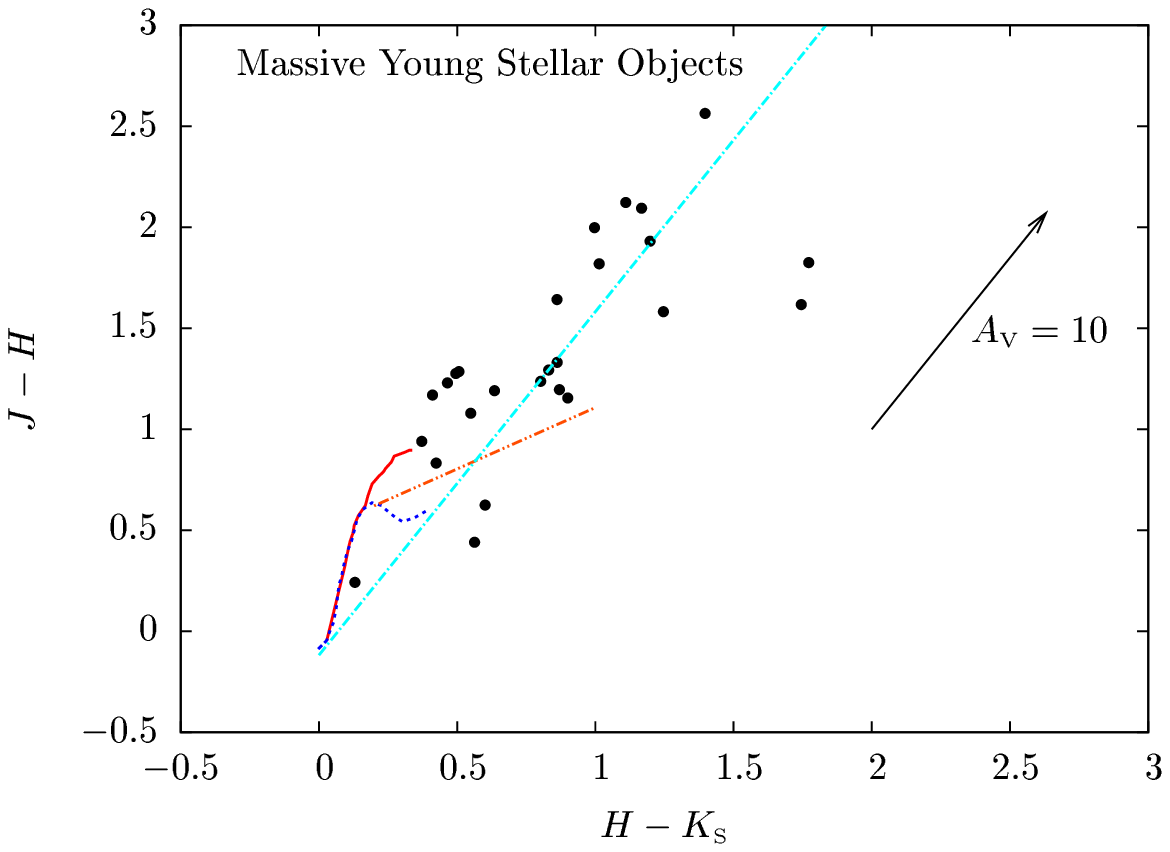}}
	\end{center}
	\end{minipage}\end{tabular}
		\caption{The distribution of four types of objects: Microquasars (upper left), 
					Cataclysmic variables (upper right), Low Mass X-ray Binaries (lower left), 
					and Massive Young Stellar Objects (lower right). 
					The stellar locus and the reddening vector are same as in Fig. \ref{CCD1}. 
  		\label{CCD2}}
\end{figure*}

\begin{table}[tbp]
\begin{center}
\begin{tabular}{rrrrr}
\hline \hline
& Microquasars & CV & LMXB & MYSO \\ \hline
I & 16 (84) & 245 (75) & 11 (73) & 27 (93) \\
II & 3 (16) & 82 (25) & 4 (27) & 2 (7) \\
total & 19 & 327 & 15 & 29 \\ \hline
\end{tabular}
\caption{
The number and percentage of objects distributed in each region. 
The values in parentheses represent percentage. 
 \label{ratio-four-objects}}
\end{center}
\end{table}

Although the locus of AGNs in the near-infrared CCD is different from that of normal stars, 
other types of objects might be distributed in the locus 
with similar properties to AGNs. 
If a position in the CCD depends on the radiation mechanism, 
other objects with radiation mechanism similar to AGNs are also expected to be located at the same position. 
Below, we further examine the loci of four types of objects which have non-thermal radiation or 
which are considered to be bright in both near-infrared and X-ray wavelengths: 
Microquasars, Cataclysmic Variables (CVs), Low Mass X-ray Binaries (LMXBs), and 
Massive Young Stellar Objects (MYSOs). 

Sample objects are extracted from three catalogues, namely 
Microquasar Candidates \citep[Microquasars; ][]{Combi2008-AA}
Cataclysmic Binaries, LMXBs, and related objects \citep[CVs and LMXBs; ][]{Ritter2003-AA}, 
and Catalogue of massive young stellar objects \citep[MYSOs; ][]{Chan1996-AAS}. 
First, we cross-identified each catalogue with 2MASS PSC, and 
extracted the near-infrared counterparts. 
\citet{Combi2008-AA} have cross-identified their catalogue with the 2MASS catalogue
 by adopting a cross-identification of $4''$. 
The positional accuracy in Ritter \& Kolb catalogue are $\sim 1''$ \citep{Ritter2003-catalogue}. 
The objects in the MYSO catalogue were selected from the Infrared Astronomical Satellite (IRAS) PSC 
whose typical position uncertainties are about $2''$ to $6''$ \citep{Beichman1988-IRAS,Helou1988-IRAS}. 
Therefore, we set positional criteria for the cross-identification 
to $\leq 2''$ (CV and LMXB catalogues) and $\leq 4''$ (Microquasar and MYSO catalogues). 
We used objects with a 2MASS photometric quality better than B (i.e., S/N $> 7\sigma$). 
Using 2MASS magnitude, they were plotted in a ($H-K_\textnormal{\tiny S}$)-($J-H$) diagram. 

Figure \ref{CCD2} shows the CCD of each object. 
In every case, a few objects are distributed around the locus of the AGNs, 
although most of the objects are distributed around the stellar locus  
or reddened region from the stellar locus. 
Table \ref{ratio-four-objects} lists the number and percentage of objects distributed in each region. 
Although the ratios of CVs and LMXBs residing in Region II are relatively larger than the other two types of objects, 
it is not more than $\sim 25\%$. 
In addition, few objects have $(H-K_\textnormal{\tiny S}) \sim 0.8$ in Region II
though most quasars/AGNs have this colour (see Fig. \ref{CCD1}). 
Accordingly, contamination by these four types of objects should be a small fraction. 

This means that the dominant radiation of the four objects should be thermal radiation. 
The AGNs also radiate thermal radiation, 
but it is a very small fraction compared with the non-thermal component produced 
by accretion around supermassive black holes. 
Therefore, AGNs should be well separated by these four objects using the near-infrared colours. 

\subsection{Contamination by normal galaxies}

\begin{figure}[thbp]
	\begin{center}
		\resizebox{90mm}{!}{\includegraphics[clip]{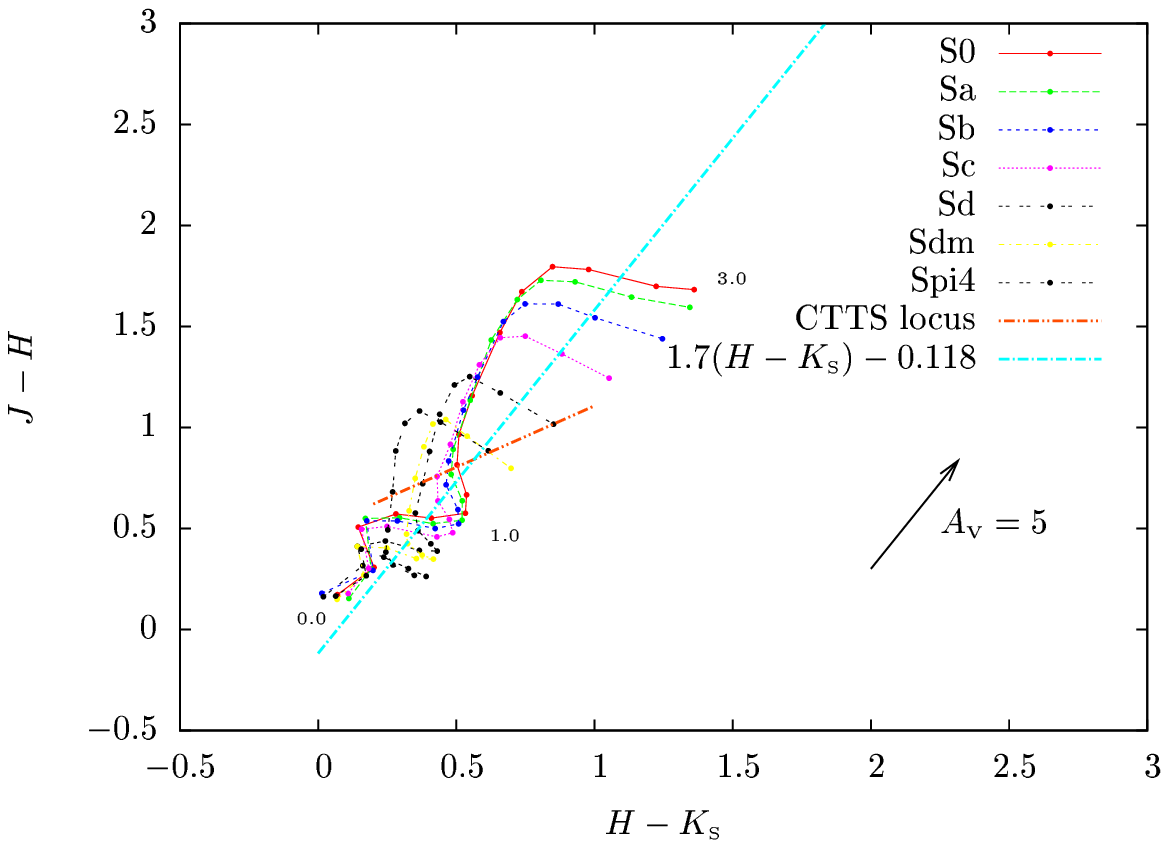}}
	\end{center}
		\caption{
		Simulated colour evolutions for seven spiral galaxies. 
				Redshift ranges from 0.0 to 3.0 with $\Delta z=0.2$ interval points. 
				The boundary between Region I and II is also drawn in the diagram. 
		\label{Simulate-galaxy-colour}}
\end{figure}

Distant galaxies that appear as point-like sources might
 contaminate the AGN locus in the near-infrared CCD. 
We confirmed the locus of normal galaxies in the near-infrared CCD 
by performing a Monte-Carlo simulation as in Sect. \ref{Simulation}. 
The SED templates we used are seven spiral galaxies in \citet{Polletta2007-ApJ}: 
spirals range from early to late types (S0-Sd).

Figure \ref{Simulate-galaxy-colour} shows the simulated intrinsic colours 
(i.e., $A_\textnormal{\tiny V}=0$) of the seven galaxies. 
Galaxies with $0\leq z\lesssim 0.8$ have intrinsic colours similar to those of normal stars 
(i.e., they are in Region I). 
Galaxies with $1.4 \lesssim z \leq 3$ are distributed
 around the reddened region of either normal stars and/or CTTS. 
Therefore, they should not be mistaken for AGN candidates. 
On the other hand, simulated colours with $z \sim 1$ are located in Region II. 
A fraction of AGN in Region II are possibly mistaken for galaxies with $z\sim 1$. 
However, galaxies at $z\sim 1$ should not have enough brightness to be detected with mid-scale telescopes. 
Even the brightest galaxy has no more than $M \sim -23$ mag at SDSS r-band \citep{Blanton2001-AJ,Baldry2004-ApJ}.
If such a galaxy were located at $z\sim 1$, 
its apparent magnitude would be $m \gtrsim 20$ mag at J-band. 
In addition, the apparent magnitude would be even fainter 
because most galaxies have $M>-23$ mag and the apparent brightness suffers extinction. 
Accordingly, only large-scale telescopes can observe these galaxies. 
Hence, few galaxies should contaminate the AGN locus in the near-infrared CCD with respect to 
the data where limiting magnitude is below 20 mag.

\section{Summary and Conclusion}
We confirmed the loci of catalogued quasars/AGNs in a ($H-K_\textnormal{\tiny S}$)-($J-H$) diagram, 
of which over $70 \sim 80\%$ are clearly separated from the stellar locus. 
In addition, 
we simulated the near-infrared colours of quasars on the basis of a Monte-Carlo simulation with Hyperz code, 
and demonstrated that the simulated colours can reproduce both the redshift-colour relations and 
the locus of quasars in the near-infrared CCD. 
We also predicted the colour evolution with respect to redshift (up to $z \sim 11$). 
Finally, we discussed the possibility of contamination by other types of objects. 
The locus of AGNs is also different from those of 
the other four probable types of objects (namely, Microquasars, CVs, LMXBs, MYSOs) 
that are expected to be located at a similar locus. 
We also demonstrated by a Monte-Carlo simulation 
that normal galaxies are unlikely to contaminate the locus of AGNs in the near-infrared CCD.

\citet{Hewett2006-MNRAS} investigated near-infrared colours of quasars using an artificial SED, 
but we proposed near-infrared colour selection criteria for extracting AGNs and 
studied both observed and simulated colours with quantitative discussions. 
An important point is that our selection criteria require only near-infrared photometric data, 
although some previous studies \citep[e.g., ][]{Glikman2007-ApJ,Glikman2008-AJ} used colour selections 
based on colours between near-infrared and optical wavelengths. 
In other words, our selection criteria make extraction of candidates easier 
because only near-infrared colours are needed. 
This technique should also be useful when we search for high-redshift quasars, 
since they become very faint in the optical wavelength due to the shift of Lyman break.

This paper demonstrates that near-infrared colours can be useful to select AGN candidates. 
If an additional constraint is imposed, more reliable candidates can be extracted. 
When we use the near-infrared colour selection with an additional constraint 
for near-infrared catalogues containing sources distributed in large area 
(e.g., 2MASS, DENIS, UKIDSS, and future surveys), 
a lot of AGN samples (possibly over $\sim$10 000) are expected to be derived in a region over $\sim$10 000 deg$^2$. \citet{Kouzuma2009-prep} \citep[see also][]{Kouzuma2009-ASPC} cross-identified between the 2MASS and ROSAT catalogues, and 
extracted AGN candidates in the entire sky using the near-infrared colour selection in this paper. 
These large number of samples may provide us with clues about such as the evolution of AGNs and X-ray background. 
Additionally, in our simulation, quasars with $z \gtrsim 8$ can be extracted on the basis of near-infrared colours. 
This property might be helpful to search for high-redshift quasars in the future.


\begin{acknowledgements}
This publication makes use of data products from the Two Micron All Sky Survey, 
which is a joint project of the University of Massachusetts and 
the Infrared Processing and Analysis Center/California Institute of Technology, 
funded by the National Aeronautics and Space Administration and the National Science Foundation.
Funding for the SDSS and SDSS-II has been provided by the Alfred P. Sloan Foundation, 
the Participating Institutions, the National Science Foundation, the U.S. Department of Energy, 
the National Aeronautics and Space Administration, the Japanese Monbukagakusho, 
the Max Planck Society, and the Higher Education Funding Council for England. 
The SDSS Web Site is http://www.sdss.org/.
We thank the anonymous referee for useful comments to improve this paper. 
\end{acknowledgements}

\bibliographystyle{aa}
\bibliography{../../JabRef/REF/all-reference}

\end{document}